\documentclass[usegraphicx,usenatbib]{mn2e}

\usepackage{amssymb,bm,color,charter}

\newcommand{\e}[1]{\times 10^{#1}}                        

\newcommand{\fig}[1]{Fig. \ref{#1}}

\pdfoutput=1

\font\tenbg=cmmib10 at 10pt
\def \rvecmu{{\hbox{\tenbg\char'026}}}

\title[Conical Winds]
{Conical Winds from the Disk-Magnetosphere Boundary}

\author[M. M. Romanova et al.]
{M. M. Romanova,$^1$\thanks{E-mail: romanova@astro.cornell.edu}, G.
V. Ustyugova$^2$\thanks{E-mail: ustyugg@rambler.ru}, A. V.
Koldoba$^3$\thanks{E-mail: koldoba@rambler.ru},  R. V. E.
Lovelace$^{1,4}$\thanks{E-mail:lovelace@astro.cornell.edu}\\
$^1$ Department of Astronomy, Cornell University, Ithaca, NY 14853, USA\\
$^2$ Keldysh Institute of Applied Mathematics, Russian Academy of
Sciences, Moscow, Russia \\
$^3$ Institute of Mathematical Modeling, Russian Academy of
Sciences, Moscow, Russia\\
$^4$ Department of Applied and Eng. Phys., Cornell University,
Ithaca, NY 14853}

\begin{document}
\maketitle \label{firstpage}

\begin{abstract}
  A new type of wind - a conical wind - has been discovered in axisymmetric magnetohydrodynamic
simulations of the disk-magnetosphere interaction in cases where the magnetic field of the star is bunched into an X-type configuration. Such a configuration arises
if the effective viscosity of the disk is larger than
the effective diffusivity, or if the
accretion rate in the disk is enhanced.
  Conical outflows flow from
the inner edge of the disk  into
a narrow shell with half-opening angle of
$30-45^\circ$.
   The outflow carries $\sim 0.1-0.3$ of the disk mass
accretion rate and part of
the disk's angular momentum.
    The conical winds are driven by the gradient of the magnetic pressure which exists above
the disk due to the winding of the stellar magnetic field.
 Exploratory 3D simulations show that conical winds are symmetric about rotation axis of
 the disk even if the magnetic dipole is significantly misaligned with the
disk's rotation axis.
    Conical winds appear around stars of different periods.
    However, in the case of a star in the ``propeller'' regime, an additional - much faster
    component appears:
an axial jet, where matter is accelerated up to very high velocities
at small distances from the star by magnetic pressure force above
the surface of the star.  The simulations are done in dimensionless units and are applicable to a variety of the disk-accreting magnetized stars: young stars, white dwarfs, neutron stars, and possibly black holes.
For the case of young stars, conical winds and axial jets may appear in different cases,
including Class I young stars, classical T Tauri stars, and EXors.
In EXors periods of enhanced accretion may lead to the formation of conical winds which correspond
to the outflows observed from these stars.

\end{abstract}


\section{Introduction}

Jets and winds are observed in many objects from young stars to systems with white dwarfs, neutron stars, and black holes (e.g., Livio 1997).
For example,  jets have recently been observed from the accreting neutron star
Circinus X-1 (Heinz et al. 2007).
  The jet shows a conical
structure with a half-opening angle about $30^\circ$. This conical
structure could be explained by precession of the jet.
However, $10$ years of observations show that orientation of the jet has not changed (Tudose et al. 2008).
Jets and outflows are also observed from symbiotic stars where matter flows from the vicinity of white dwarfs during periods of enhanced accretion (e.g., Sokoloski \& Kenyon 2003). It is often the case that outflows are associated with enhanced accretion
(e.g., Cabrit et al. 1990; Lovelace, Romanova, Newman 1994).

A large body of observations has been accumulated on outflows
form young stars at different stages of their evolution from very young stars where powerful jets are observed up to classical T Tauri stars (CTTSs) where
weaker jets and winds are observed
 (see review by Ray et al. 2007).
    A significant number of CTTS
show signs of outflows in spectral lines, in particular in
He I (Edwards et al. 2006; Kwan, Edwards, \& Fischer 2007).
High-resolution observations show that outflows often have an
``onion-skin" structure, with high-velocity outflows in the axial
region, and lower-velocity outflow at larger distance from the axis
(Bacciotti et al. 2000).
   High angular resolution spectra of
[FeII]$\lambda~ 1.644\mu$m emission line taken along the jets of
DG Tau, HL Tau and RW Auriga reveal two components: a
high-velocity well-collimated extended component with
velocity $v\sim 200-400$ km/s and a low-velocity $v\sim 100$ km/s uncollimated component which is close to the star (Pyo et al. 2003, 2006).
High-resolution observations of molecular hydrogen in HL Tau have shown that at small distances from the star the flow shows a conical structure with outflow velocity $\sim 50-80$ km/s (Takami et al. 2007).
EXors represent an interesting stage of evolution of young stars where enhanced accretion leads to period of enhanced outflows
(e.g., Brittain et al. 2007)

 Different models have been
proposed to explain outflows from CTTSs (see review by Ferreira,
Dougados, \& Cabrit 2006).
  The models include those where the outflow originates from a radially distributed disk wind (K\"onigl \& Pudritz 2000; Ferreira et al. 2006) or  from the innermost region of the accretion
disk (e.g., Lovelace, Berk \& Contopoulos 1991; Livio 1997).
   Further, there is the
X-wind  model (Shu et al. 1994; 2007; Najita \&
Shu 1994; Cai et al. 2008) where most of the matter outflows from the disk-magnetosphere boundary.
    This work focuses on the outflows from the inner
part of the disk or disk-magnetosphere boundary.
    We discovered
conditions which favor conical outflows and performed axisymmetric and exploratory 3D MHD simulations for both
slowly and rapidly rotating stars including stars in the propeller
regime.

Earlier we investigated  outflows from  accreting
rapidly -rotating magnetized stars in the ``propeller"  regime (e.g., Illarionov \& Sunyaev 1975; Lovelace, Romanova \&
Bisnovatyi-Kogan 1999; Alpar \& Shaham 1985) where the magnetosphere rotates faster than the inner regions of the disk and the magnetosphere transfers its
angular momentum to the disk matter (e.g. Lovelace et al. 1999).
Simulations show that most of matter flows from the inner parts of
the disk to wide-angle open cones.  At the same
time  most of the angular momentum and rotational energy of the star  outflows along open stellar magnetic field lines in a
collimated, magnetically dominated jet of low-density and high-velocity (Romanova et al. 2005; Ustyugova et al. 2006).

In the present study we were able to obtain conical winds
in cases of $\it slowly$ rotating stars, where  new disk matter arrives after period of lower accretion. In this situation the
poloidal field lines are bunched into an X-type
configuration and persistent conical winds are observed. Compared to X-winds proposed earlier by Shu and collaborators (e.g. Shu et al. 1994) conical winds
are strongly non-stationary and they may originate from a star with any rotation rate.
Such winds may appear and blow during periods of enhanced accretion, or may appear in
any situation where the inward accretion velocity of the disk is higher than outward diffusion of the
magnetic field. Simulations were done in dimensionless form and such winds may appear around magnetized stars of different sizes from young stars to neutron stars, and also possibly around magnetized black holes.

In \S 2 we describe the model. In \S 3 we describe the physics of conical outflows.
In \S 4 we discuss dependence of the results on different parameters. In \S 5 we show exploratory 3D simulations of conical outflows. In \S 6 we discuss application of the model to EXors and CTTSs. In \S 7 we compare conical winds with propeller-driven winds and other types of winds. In \S 8 we present discussion and conclusions.

\begin{figure*}
\centering
\includegraphics[width=6in]{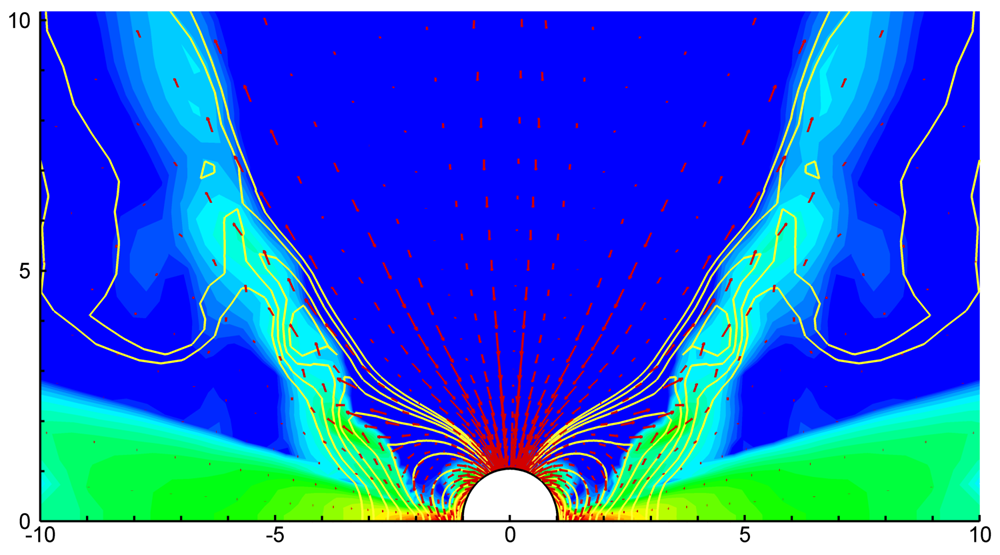}
\caption{A snapshot of the outflow to conical winds, at the moment
of time $T=500$. The background shows logarithm of the absolute value of
the poloidal matter flux,
lines are the sample magnetic field lines, and arrows are vectors of
the poloidal velocity.}\label{sym-big}
\end{figure*}

\begin{figure*}
\centering
\includegraphics[width=5in]{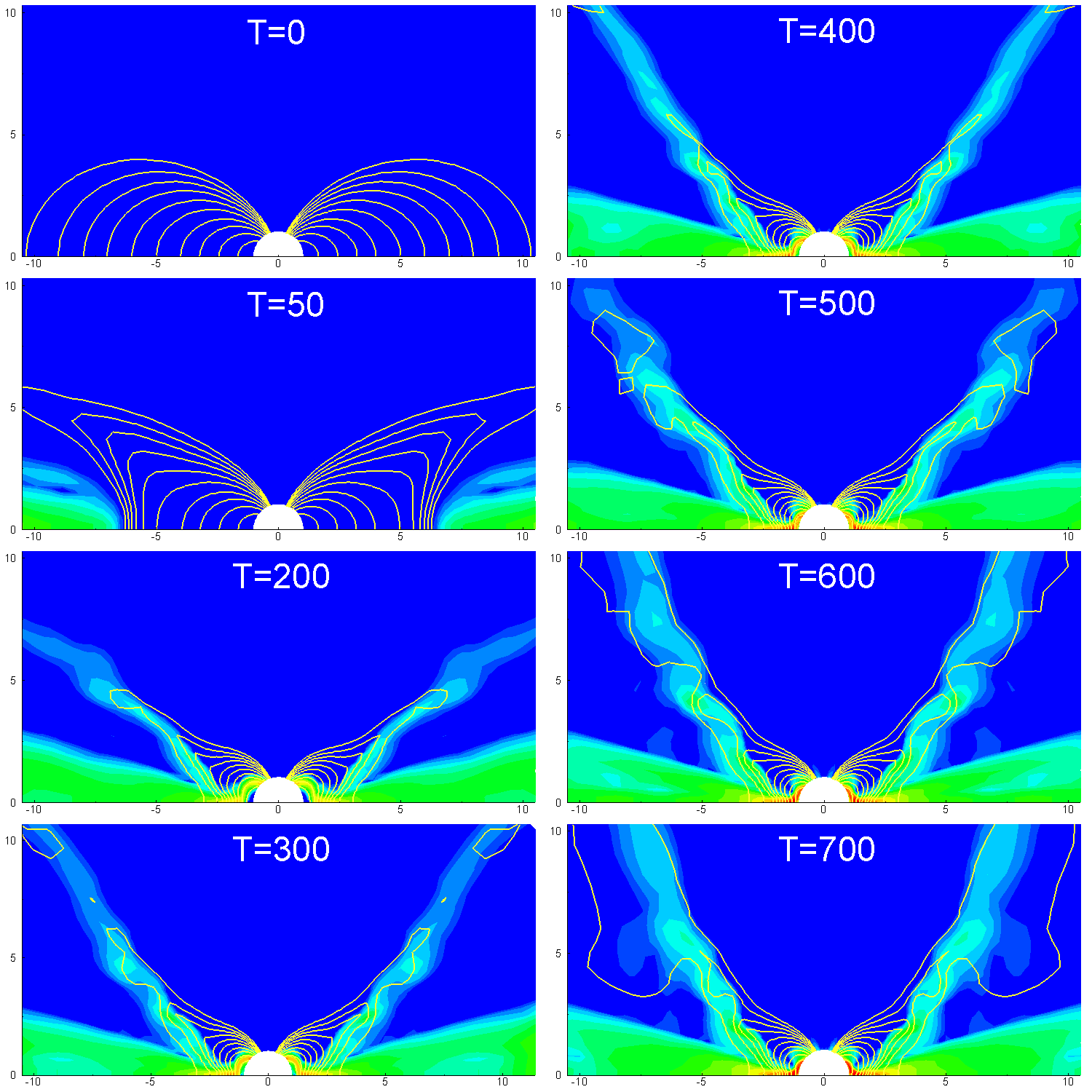}
\caption{Figure shows formation and evolution of conical winds
with time. The background shows logarithm of the absolute value of
the poloidal matter flux, lines - are selected magnetic field
lines.}\label{sym-8}
\end{figure*}

\begin{figure*}
\centering
\includegraphics[width=6in]{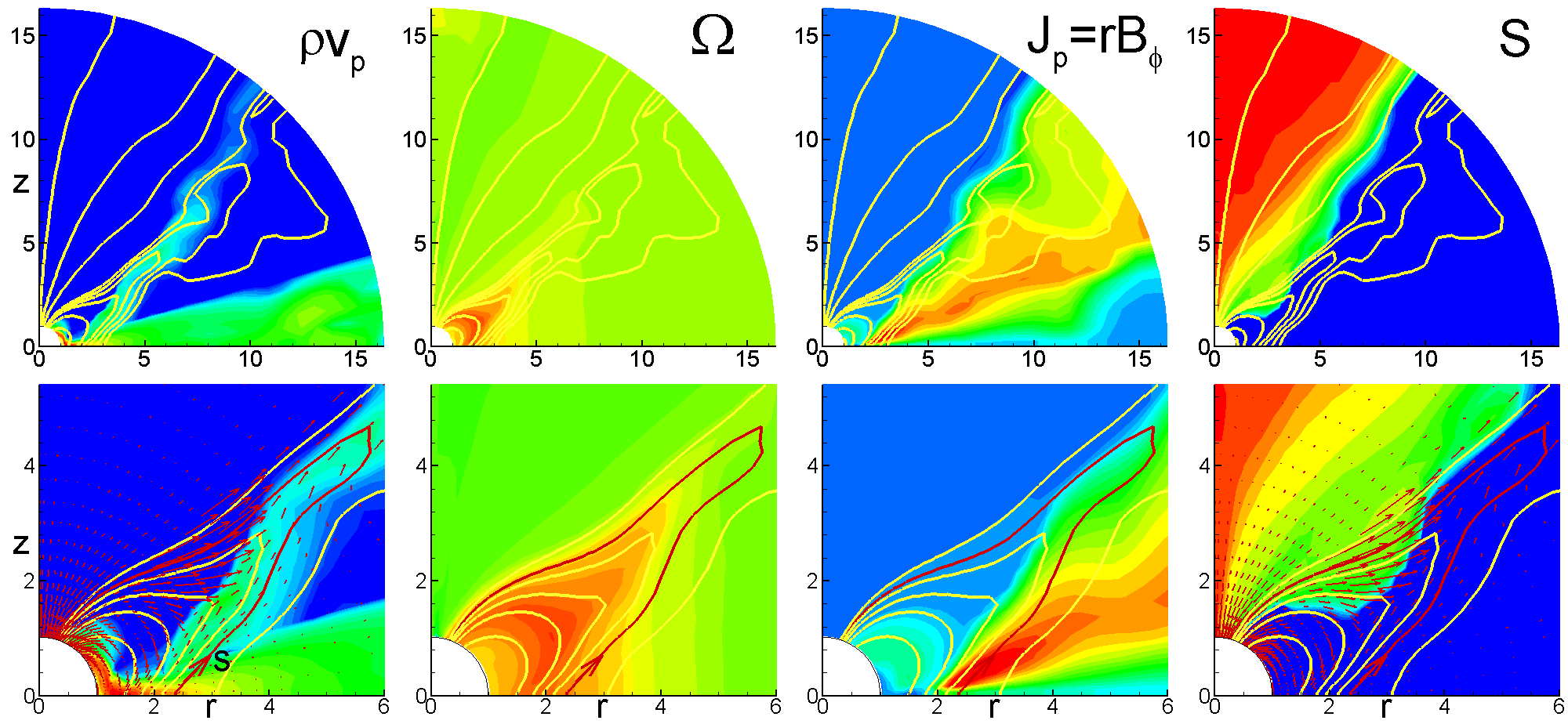}
\caption{Different values shown for one moment of time, $T=500$ in
the whole region (top panels) and near the star (bottom panels). The
background shows (from left to right): the logarithm of the absolute
value of the poloidal matter flux, $\rho
v_p$, angular velocity, $\Omega$, poloidal current, $J_p=r B_\phi$,
and entropy $S$. Lines are selected magnetic field lines. Thick line
is a magnetic field line which is used for analysis of flow in the
direction of conical winds. Vectors are velocity vectors. An arrow
shows that the distance s along the line used in the subsequent
plots starts at the disk.}\label{8-all}
\end{figure*}

\begin{figure*}
\centering
\includegraphics[width=5in]{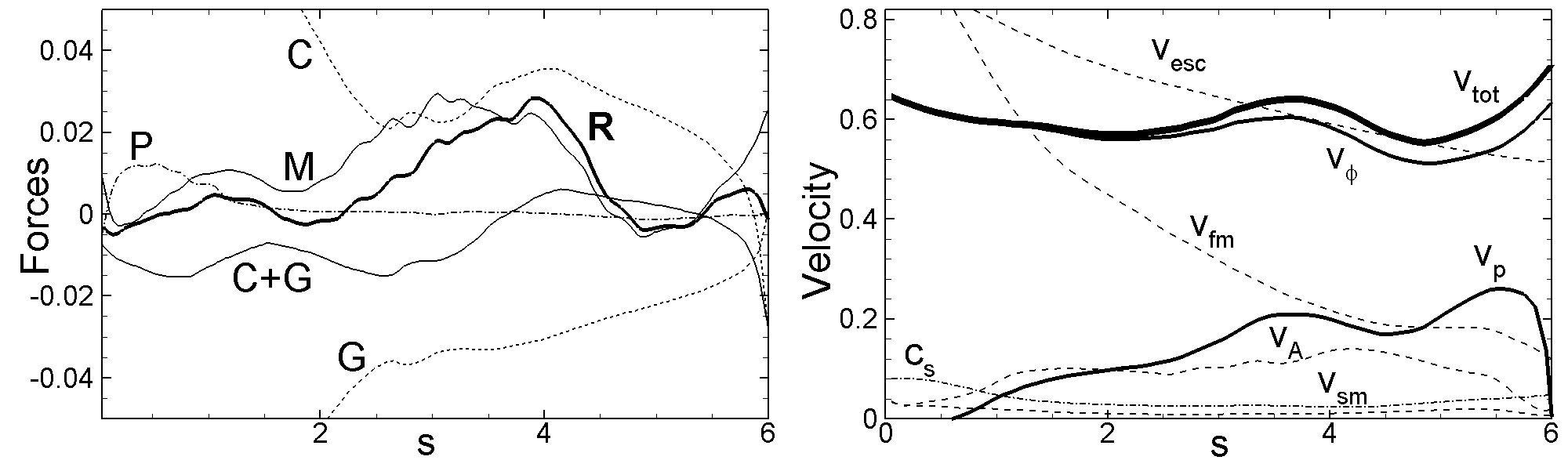}
\caption{Left panel: Forces along the part of the magnetic field
line shown on the \fig{8-all} starting from the disk up to
the point when the line curves towards the star. Labels are:
G-gravitational, C-centrifugal, M-magnetic, P-pressure gradient, and
R - resulting forces.  Right panel: velocities projected to the same
line: $v_{tot}$ - total, $v_P$ - poloidal, $v_\phi$ - azimuthal, $c_s$ - sound speed,
$v_A$ - Alfv\'en, $v_{sm}$ - slow magnetosonic, $v_{fm}$ - fast
magnetosonic, $v_{esc}$ - escape velocities.}\label{force-vel}
\end{figure*}

\begin{figure*}
\centering
\includegraphics[width=6in]{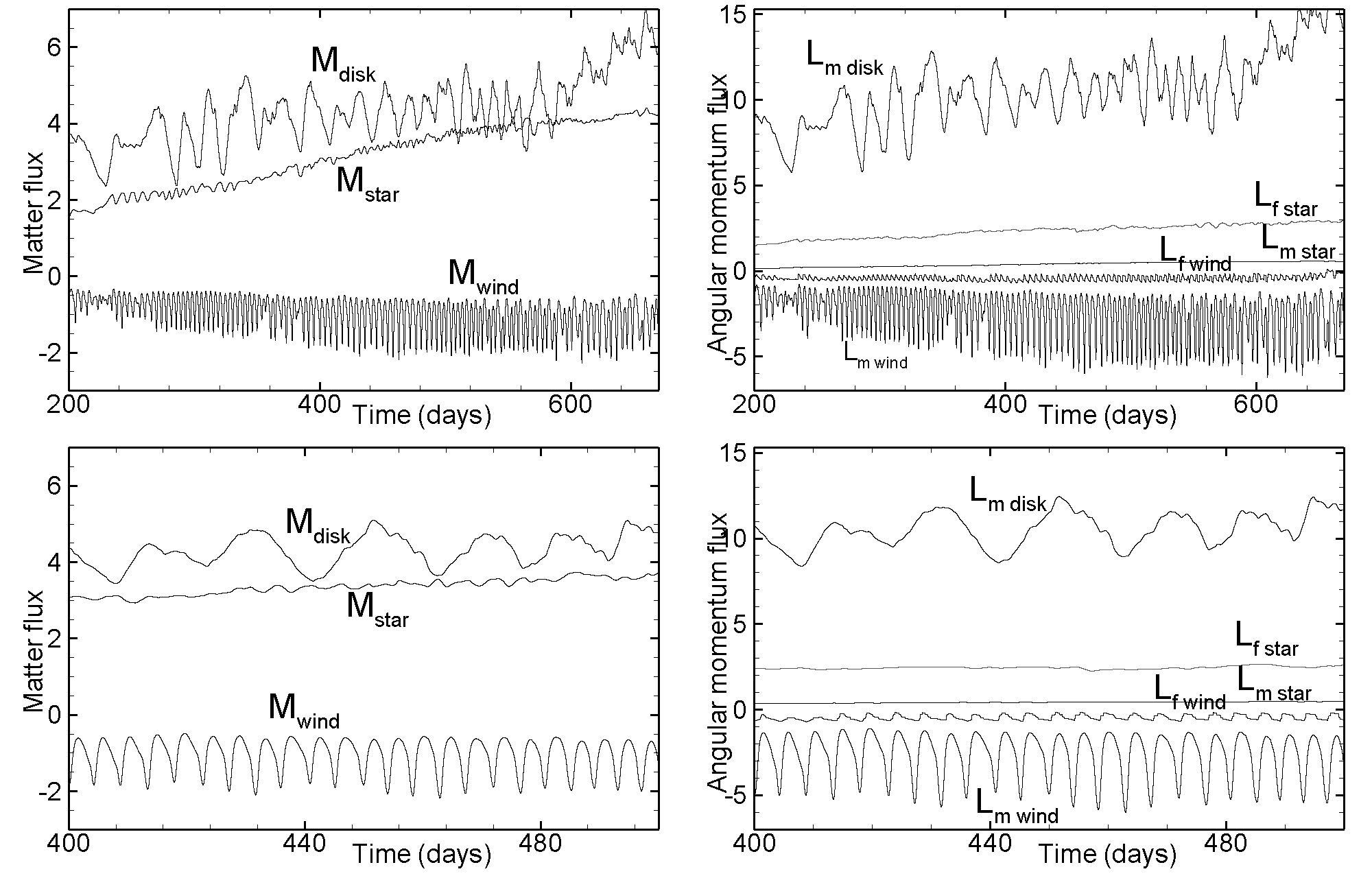}
\caption{Left panels: matter fluxes to the star $\dot M_{star}$, to
conical winds $\dot M_{wind}$ and through the disk $\dot M_{disk}$.
Right panels: angular momentum fluxes carried by the disk, $L_{disk}$, to
the star carried by matter, $L_{m star}$, and by magnetic field,
$L_{f star}$, and to conical winds, $L_{m winds}$ and $L_{f winds}$.
Bottom panels show fluxes during a part of the simulation
time.}\label{fluxes-4}
\end{figure*}

\begin{figure*}
\centering
\includegraphics[width=6in]{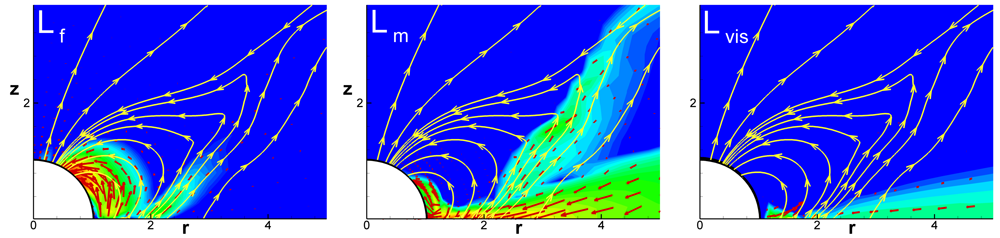}
\caption{Left panel: the background, vectors and streamlines show
angular momentum carried by magnetic field. Middle panel: background
and arrows show angular momentum carried by matter, streamlines show
angular momentum carried by the field. Right panel: the background shows angular momentum
caried by viscosity.}\label{ang-3}
\end{figure*}

\begin{figure*}
\centering
\includegraphics[width=6in]{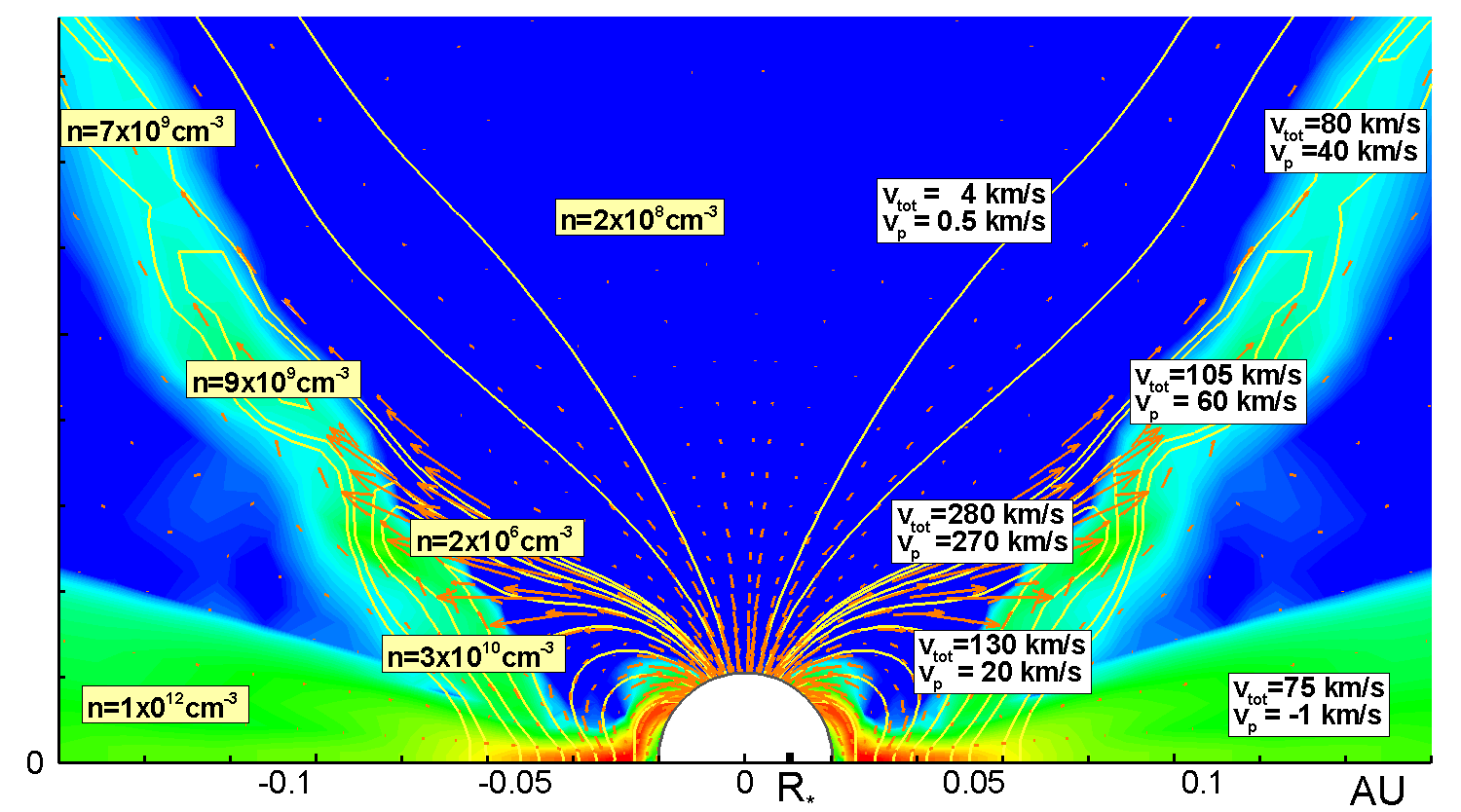}
\caption{Density and velocity distributions in conical winds for CTTS. Here we converted
dimensionless values obtained from simulations to dimensional values
for typical T Tauri star taken from the Table.}\label{con-numb}
\end{figure*}

\begin{figure*}
\centering
\includegraphics[width=6in]{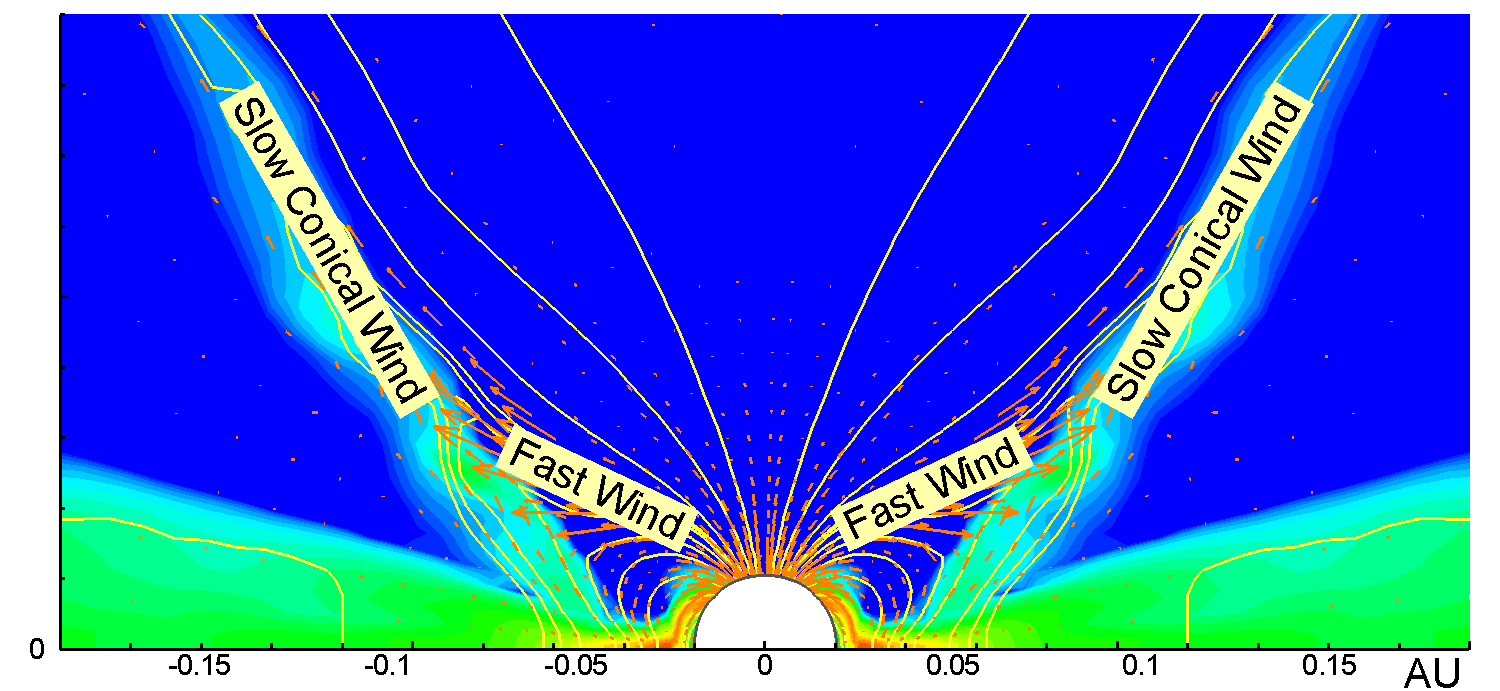}
\caption{Two components of winds from slowly rotating star are
labeled.}\label{con-label}
\end{figure*}

\begin{table}
\begin{tabular}{l@{\extracolsep{0.2em}}l@{}lll}

\hline
&                                                   & CTTSs       & White dwarfs          & Neutron stars           \\
\hline

\multicolumn{2}{l}{$M(M_\odot)$}                    & 0.8           & 1                     & 1.4                     \\
\multicolumn{2}{l}{$R$}                             & $2R_\odot$    & 5000 km               & 10 km                   \\
\multicolumn{2}{l}{$R_0$ (cm)}                      & $2.8\e{11}$   & $1.0\e9$              & $2\e6$                \\
\multicolumn{2}{l}{$v_0$ (cm s$^{-1}$)}             & $1.95\e7$     & $3.6\e8$              & $9.7\e{9}$                \\
\multicolumn{2}{l}{$\omega_0$ (s$^{-1}$)}           & $2.0\e{-4}$   & 0.36                   & $4.8\e3$                \\
\multicolumn{2}{l}{$P_0$}                           & $1.04$ days   & $17.2$ s               & $1.3$ ms               \\
\multicolumn{2}{l}{$B_{\star_0}$ (G)}               & $10^3$        & $10^6$                & $10^9$                  \\
\multicolumn{2}{l}{$B_0$ (G)}                       & 100           & $1.2\e4$                & $1.2\e{7}$               \\
\multicolumn{2}{l}{$\rho_0$ (g cm$^{-3}$)}          & $4.1\e{-13}$  & $3.7\e{-8}$           & $1.7\e{-6}$          \\
\multicolumn{2}{l}{$n_0$ (1/cm$^{-3}$)}             & $2.4\e{11}$   & $7.0\e{14}$           & $1.0\e{18}$          \\
\multicolumn{2}{l}{$\dot M_0$($M_\odot$yr$^{-1}$)}  & $2.0\e{-8}$   & $1.3\e{-8}$           & $2.0\e{-9}$          \\
\hline
\end{tabular}
\caption{Sample reference values of the dynamical quantities used in
our simulations.} \label{tab:refval}
\end{table}

\section{Model}

We have done axisymmetric and exploratory 3D MHD simulations of the interaction of an
accretion disk with the magnetosphere of a rotating star. The
corotation radius of the star $r_{cr}=(GM_*/\Omega_*^2)^{1/3}$ is
in most cases
larger than the magnetospheric radius $r_m$ which is determined by the balance between the pressure of the star's magnetic field and the ram pressure of the disk matter.

The  model we use is similar to that described in Romanova
et al. (2005) and Ustyugova et al. (2006).  Specifically, (1) a
spherical coordinate system $(r,~\theta,~\phi)$ is used which gives
high grid resolution near the dipole; (2) the complete set of MHD
equations is solved to find the eight variables $(\rho,~ v_R,~
v_\theta,~ v_\phi, ~B_r, ~B_\theta, ~B_\phi, ~\epsilon$) (with $\epsilon$
the specific internal energy); (3) a Godunov-type numerical method
is used; (4) an effective viscosity and effective magnetic diffusivity are incorporated
the code using an $\alpha-$model  controlled by dimensionless parameters $\alpha_{\rm
v}$ and $\alpha_{\rm d}$. We varied both parameters in the range
$0.01 - 1$.

\smallskip
{\bf 2.1.~ Initial and Boundary Conditions.}
 Initially we suggest that the disk is far away from the star
 (beyond the external boundary of the simulation region),
so that the non-perturbed dipole field of the star initially occupies the entire simulation region and the magnetic field of the star is a dipole: ${\bf B}=[3(\rvecmu \cdot {\bf r}) \bf r -\rvecmu {\bf r}^2]/|{\bf r}|^5$, where $\rvecmu$ is the star's
magnetic moment. The simulation region is filled with a
low-density corona and is initially non-rotating.
    The corona is isothermal and  in thermal equilibrium,
so that the density and pressure distributions are: $\rho=\rho_0
\exp[GM/(c_s^2 r)]$ and $p=c_s^2 \rho_0 \exp[GM/(c_s^2 r)]$, where $c_s$ is the isothermal sound speed, $\rho_0$ is the matter
density at the infinity (approximately equal to that on the external
boundary).  To ensure the gradual start-up (smooth initial
conditions) we start to rotate star with very small angular velocity
corresponding to the corotation radius at the external boundary
$R_{out}$, $r_{cor} = R_{out} (GM/\Omega_d^2)^{1/3}$ and later
gradually increase the angular velocity form that up to much larger values
$\Omega$ during 10 periods of  rotation of the disk at $r=1$.
The (final) angular velocity of the star was different in different runs ranging from very
small (corresponding to $r_{cor}=10$) up to very large (corresponding to $r_{cor}=1$) with most
of runs done at intermediate value $r_{cor}=3$ (angular velocity of
the star in dimensionless units $\Omega_*\approx 0.19$). Spinning-up of the star leads
to spinning-up of the magnetosphere and the low-density corona.
We checked that rotating dipole brings the corona to corotation in a very short time
(with a speed corresponding to propagation of a
torsional Alfv\'en waves through the corona).

We divided the external boundary to the region of corona, $\theta< \theta_{disk}$
and the region of the disk, $\theta> \theta_{disk}$. In the corona  we take ``free" boundary conditions for all variables, but do not
permit matter to flow back to the simulation region from the
boundary. In the disk region,  we fix the density at
some value, $\rho=\rho_d$, and establish slightly sub-Keplerian velocity, $\Omega_d=\kappa\Omega(r_d)$, where $\kappa = \sqrt{1 -
0.003}$ so that matter flows into the simulation region through the boundary. When the matter enters the simulation region it continues to
flow inward due to the viscosity incorporated in the
code. We use an $\alpha-$type viscosity with viscosity coefficient
proportional to $\alpha_{vis}$. Diffusivity is incorporated into the
code with diffusivity coefficient proportional to coefficient
$\alpha_{dif}$. Inflowing matter has zero magnetic flux.
  The boundary conditions on
the equatorial plane and on the rotation axis are symmetric and antisymmetric (see e.g. Ustyugova et al. 2006).

\smallskip
{\bf 2.2.~Reference Units}. The MHD equations were solved in
dimensionless form so that results can be applicable to stars of
different scales, such as to  classical T Tauri stars (CTTSs), to
accreting white dwarfs, and to neutron stars. We chose a reference
mass $M_0=M_*$ to be typical mass for different types of stars (see Table). The reference radius is twice the radius of the star, $R_0=2 R_*$ and the surface
magnetic field $B_*$ is different for different types of stars.
The reference velocity is $v_0=(GM_0/R_0)^{1/2}$,
the time-scale $t_0=R_0/v_0$, and the angular velocity $\Omega_0=1/t_0$. We
measure time in units of the rotational period with Keplerian speed
at $r=R_0$: $P_0=2\pi t_0$ and in the plots show dimensionless time $T=t/P_0$. Other reference values:
the reference density
$\rho_0$ and number density of particles $n_0$, the reference
accretion rate $\dot M_0=\rho_0 v_0
R_0^2$ and the reference angular momentum flux $\dot L_0=\dot M_0 v_0
R_0$. Table 1 shows examples of reference variables for
different stars.

  We solved the MHD equations for normalized variables:
$\tilde\rho=\rho/\rho_0$, $\tilde v=v/v_0$, $\tilde B= B/B_0$, etc.
The plots show the normalized variables (with tilda's
implicit).  Examples in dimensional variables are given
for young stars.

\section{Physics of Conical Winds}

Multiple runs were done in the axisymmetric case for a wide variety of parameters. We chose one set of parameters as our main
case  and describe it in detail.
We observed that the condition $\alpha_{vis} > \alpha_{dif}$ is  necessary for
conical winds to form. That is, the magnetic Prandtl number
of the turbulence,
$Pr=\alpha_{vis}/\alpha_{dif} > 1$. In addition, the conical winds
are stronger if the diffusivity is not very small, that is if
$\alpha_{dif}>0.03$. Below we show results for
$\alpha_{vis}=0.3$ and $\alpha_{dif}=0.1$ which we term ``the main case".

For the main case we chose
densities in the disk and corona:  $\tilde\rho_d=10$,
$\tilde\rho_c=0.001$, corotation radius $\tilde r_{cor}=3$.  Results can be
applied to different types of stars.  However ,as an example, we
show results in dimensional units for typical CTTS, where the
reference values are taken from Table 1. For example, for a CTTS
with a $\tilde P_*=5.4$ days, the unit of time used in figures below is $P_0=1.04$ days (see Table, CTTS column).

\subsection{Matter flow, velocities and forces}

    The simulations show that when the disk comes close to the star and bunches the poloidal field lines into an
X-type configuration then conical outflows start and continue for a long time, as long as the simulations were run. The \fig{sym-big} shows a typical view of the matter flow where the matter flux is shown as a background. Matter flows to the cone
into a relatively narrow shell and the
cone has a half-opening angle, $\theta=30^\circ-40^\circ$. The process is non-stationary due to episodes of inflation and reconnection of the magnetic field lines (see animation at
http://www.astro.cornell.edu/$\sim$romanova/conical.htm.
One can
see that the disk comes quite close to the star. Test simulations (which require much longer runs) have been done for an
inner boundary smaller by a factor of two. This showed that
the conical winds are similar, while the magnetosphere is larger.
The poloidal velocity vectors show that matter is accelerated and
forms a hollow conical wind.  At the same time low-density matter is
accelerated up to high velocities (hundreds of km/s in CTTSs case) along the inner surface of the cone (i.e., closer to the axis).
This high-velocity component may be important in explanation of some highly blue-shifted spectral lines
which form near CTTSs (Edward et al. 2003; 2006).
  Matter which is accelerated in this region
may come from the star and/or be captured from the main accretion flow.

The \fig{sym-8} shows that the conical winds form and later become self-supported for a long time. Figure shows that all magnetic flux which threaded the disk at $T=0$ is bunched
later by the disk into X-type configuration. The inward motion of the disk is slow: it took about $T=100$ rotational periods $\tilde P_0$ (at $r=1$) to reach the vicinity of the star. The outflows started at time $T\approx 150$ and continued for
many hundreds of rotations.
The magnetic field in the
conical winds reconnects frequently and matter is expelled to
conical winds in blobs with time-scale about $(5-6) \tilde
P_0$ (see animation at http://www.astro.cornell.edu/$\sim$romanova/conical.htm ).

The \fig{sym-big} shows conical outflows at one  moment of time
$T=500$. The figure also shows velocity vectors of the flow. Most of the matter is accelerated along conical winds. Some low-density matter is accelerated to much higher velocities on the inner
side of the cone.

The \fig{8-all} shows the distribution of different parameters as a
background at  time $T=500$ in the whole region (top)
and in the inner part of the region (bottom). The middle panels show the distribution of angular velocity $\Omega$ and lines of magnetic flux. One can see that the inner region of the closed magnetosphere
rotates with the angular velocity of the star ($1.2 < r < 2$),
while the outer region of the corona above the disk (at $r > 2$) rotates with the
angular velocity of the disk. The field lines which start
at the disk go through the regions of decreasing angular velocity and thus they are strongly wound up owing to
the difference in the angular rotation rates.
 This leads to strong poloidal current $r B_\phi$
above the disk (see next panels to the right) and the
magnetic force associated with gradient of
magnetic pressure, $F_m\sim  -grad[(r B_\phi)^2]$ appears. This is the main force driving matter to conical winds (as proposed
by Lovelace et al. 1991).
The right-hand panels (with entropy as a background) show that matter flowing to conical winds is cold.

The \fig{force-vel} shows the projection of the different forces onto a field
line (shown in bold on the bottom panels of \fig{8-all}).
This is one of the field lines which crosses the region
where matter is accelerated to conical winds so that we calculate
projection of forces to this line starting from the disk and
approximately half-way to the star (before it bends towards the
star). The \fig{force-vel} shows that
both, centrifugal $C$ and gravitational $G$ forces are large and
approximately compensate each other. The centrifugal force is
smaller, because a star rotates slowly, so that the sum $G+C$ is
negative. The pressure gradient force $P$ is positive inside and near
the disk, but it is only strong enough to compensate negative value
of $G+C$. The magnetic force $M$ is the main one
acting in the positive direction and accelerating matter to conical
winds. The gradient of this force is directed approximately in the $z-$ direction,
and matter is pushed vertically up. However, it is stopped by the
magnetic flux of the dipole and is redirected into a cone. This is
why the hollow conical flow is relatively thin and has a relatively small half-opening angle. If the centrifugal force were dominate (as in X-winds,
Shu et al. 1994) , then the cone would have larger opening angle,
and it would not be so narrow, because the centrifugal force has
horizontal direction. In addition, the vertical direction of the
magnetic force leads to frequent forced reconnection of the magnetic flux threading the conical flows.

The right-hand panels of \fig{force-vel} show the velocity variation along the same same magnetic field line.
One can see that the total velocity $v_{tot}$ is almost solely determined by the azimuthal rotation of the flow in the conical wind. Matter is accelerated in the poloidal direction
from $\sim c_s$ (near the disk)up to slow magnetosonic $v_{sm}$,
Alfv\'en $v_A$
and fast magnetosonic $v_{fm}$ velocities. The azimuthal
component of velocity $v_\phi$ is much larger than poloidal
component, $v_\phi > 3 v_p$. The projection of the poloidal velocity
to the chosen field line decreases at larger distances $s$  because the field
line curves towards the star. In reality, matter is accelerated
more, up to $\sim v\approx 0.5$ in dimensionless units. Note that the sound speed $c_s$ is small so that the pressure
gradient is not significant in driving the conical winds.

\subsection{Matter and angular momentum fluxes}

We calculated the matter fluxes to the star's
surface (integrated over
 $R=1$), to the conical wind
(calculated at $R=6$), and the matter flux through the
disk at $R=6$. The \fig{fluxes-4} (left panels) show matter
fluxes. One can see that most of matter incoming in
the disk $\dot M_{disk}$ accretes to the star, $\dot M_{star}$,
while a smaller part ($1/3$ in the main case) flows to the conical
wind.  The bottom panel shows matter flux curves for a small interval of
time for better temporal resolution. We also calculated the angular
momentum fluxes  to the star, to the conical wind, and through the disk.

 We calculated the flux of angular momentum through the surface of the star
 and through the surface $R=6$. In both cases
 the flux was calculated as
$$
\dot{L} = \int d {\bf S}\cdot r \sin \theta \left[ \rho v_\phi {\bf v}_p
- \frac{B_\phi {\bf B}_p}{4 \pi}
- \nu_t \rho r \sin \theta \nabla \omega \right]~,
$$
where $d {\bf S}$ is the surface element directed outward to the
region. The first term on the right-hand-side ${\bf L}$
gives  the transport of angular momentum by the matter;
the second term is magnetic field
contribution; and the third term is the transport due to
viscous stress. We also calculated the density of the angular momentum flux
and show it as a background in the \fig{ang-3}.

The \fig{fluxes-4} (right panels) shows that the largest flux is
the inward flux carried by the disk. Only small part ($\approx 1/3$)
of this flux is carried away by conical winds. Another small part
($\approx 1/4$)  is transported to the (slowly rotating) star mainly
through magnetic interaction with much smaller portion transported
by the matter. The rest of the flux is transported back by viscous
stresses.

The \fig{ang-3} shows angular momentum flow. The left panel shows that angular momentum
to the star is transported by magnetic field. Most of transport
occurs through magnetic field lines in the area of the funnel
stream, where most of matter flows (see also Romanova et al. 2002;
Bessolaz et al. 2007). One can see that the magnetic stress is also high at
the base of the conical outflows which means that it is responsible for transport of the angular momentum at the base of the conical winds. Middle panel shows that
later, above the disk, angular momentum is carried by matter. Right panel shows angular momentum carried by viscous stress.

\subsection{Application to CTTSs}

\begin{figure}
\centering
\includegraphics[width=3in]{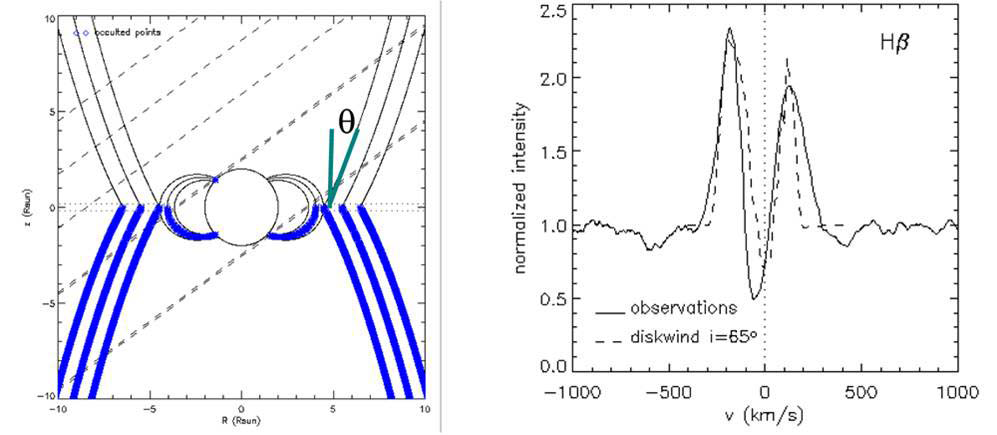}
\caption{Modeling of the  $H_\beta$ line in RW Aurigae led to the
conclusion that a conical shaped wind with half-opening angle
$30-40^\circ$ and a narrow annulus gives the best match to the
observations of this line (from Alencar et al. 2005).}\label{rw_aur}
\end{figure}

CTTSs are strongly variable on different time-scales
including a multi-year scale (Herbst et al. 2004; Grankin et al.
2007). This is connected with variation of the accretion rate
through the disk which may lead to the enhancement of outflows
(e.g., Cabrit et al. 1990). Simulations have shown that the bunching
of field lines by the new matter after period of the low
accretion may lead to quite long outburst of matter to  conical
winds and may be the reason for formation of micro-jets in the
CTTSs. If CTTS is in a binary system, then an accretion rate may be
episodically enhanced due to interaction with the secondary star.
Events of fast, implosive accretion are possible due to thermal
instability or global magnetic instability, where the accretion rate
is enhanced due to the formation of disk winds (Lovelace et al. 1994). \fig{con-numb} shows an example of conical winds in case of CTTSs.
Some observations confirm a hollow cone shape of the
outflows from CTTS. For example comparison of possible geometries of outflows observed in  $H_\beta$ line in RW Aurigae led to the
conclusion that a conical shaped wind with half-opening angle
$30-40^\circ$ and a narrow annulus gives the best match to the
observations of this line (from Alencar et al. 2005; see \fig{rw_aur}).

Two components of the outflow are observed (see \fig{con-label}) -
a slow component, associated with the conical wind, and a fast component associated with accelerated  low-density matter closer
to the rotating axis. This high-velocity component may be important in understanding some highly blue-shifted spectral lines
which form near CTTSs. Matter which is accelerated in this region
may come from the star or it may be captured from the main
accretion flow. Observations show that a significant number of CTTS
show signs of outflows in spectral lines, in particular in
He I (Edwards et al. 2006; Kwan, Edwards, \& Fischer 2007).

If CTTS typically  in rotational equilibrium state (e.g. Long et al. 2005), then the stars will oscillate between
periods of accretion and periods of propeller-driven
outflows. In the propeller stage the high velocity component
of the outflow is even stronger (see \S 5).

\subsection{Periods of enhanced accretion and outflows in EXors}

\begin{figure*}
\centering
\includegraphics[width=6in]{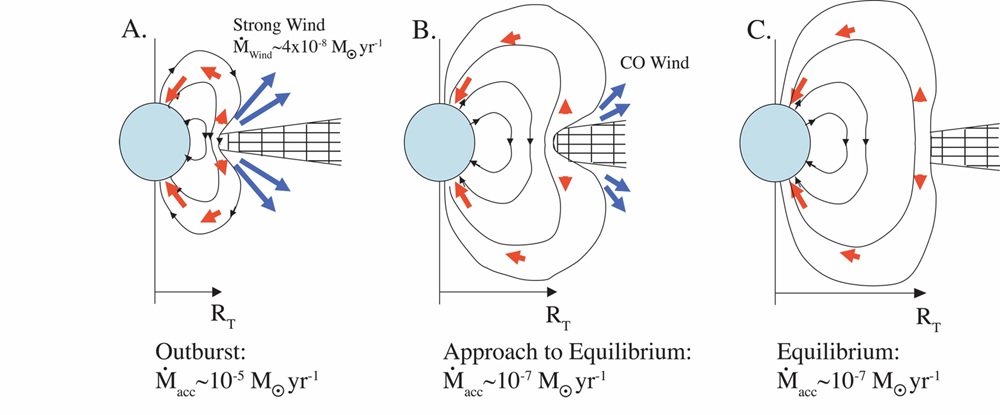}
\caption{Schematic model of an Exor V1647 Ori. During the outburst
the accretion rate is enhanced so that the magnetospheric radius
$R_m$ decreases and the magnetic field lines were bunched (A). This
results in a fast, hot outflow. As the accretion rate decreases, the
disk moves outward and this results in a slower, cooler CO outflow
(B). Further decrease in the accretion rate leads to a quiescence
state where the production of warm outflows stops (C). From Brittain
et al. (2007)}\label{brittain}
\end{figure*}

A period of enhanced accretion may lead to outbursts in EXors,
where the accretion rate increases up to $10^{-6} - 10^{-5} M_{\odot}/$/yr and
strong outflows are observed. Brittain et al. (2007) reported on the
outflow of warm gas  from the inner disk around EXor V1647 observed
in the blue absorption of the CO line during the decline of the EXor
activity.
   He concluded that this outflow is a continuation of
activity associated with early enhanced accretion and bunching of
magnetic field lines (see \fig{brittain}).

Our simulations are directly applicable to such a situation: new matter comes
after period of low accretion rate in CTTS and  magnetic field lines are bunched close to the star. The accretion rates are higher than in CTTSs so that all dimensional
parameters will correspond  to stronger outflows.
In our main example of a CTTS, the disk
stops at $R_m=2.4R_*$. In EXors, we take the radius of a star at the
\fig{con-numb} equal to the inner boundary, so that the disk stops much
closer to the star, $R_m=1.2 R_*$. Then all velocities are a factor
$1.4$ higher and densities a factor of $32$ higher (compared to \fig{con-numb}), and the matter flux in \fig{fluxes-4} is a factor of $11$ higher
than in the main example relevant to CTTSs.

\subsection{Comparison of the model with outflows from the
Circinus X-1 - the neutron star hosting binary}

Circinus X-1 represents a case where jets are seen from the vicinity
of the accreting neutron star. The system is unusual in the sense
that Type I X-ray  bursts as well as twin-peak X-ray QPOs are observed.   The neutron star is estimated to have a
weak magnetic field (Boutloukos  et al. 2006).
    The binary system in this case has a high
eccentricity ($e\sim 0.4-0.9$)  and thus has periods of low and high
accretion rates (e.g., Murdin 1980). Two-component outflows are
observed.
    The radio observations  show a
non-stationary jet on both arcmin and arcsecond scales, which
appears to have a small opening angle. Recent spectroscopic
observations in optics (Jonker et al. 2007) and in X-ray band (Iaria
et al. 2008; Schulz et al. 2008) have shown that outflows have a
conical structure with a half-opening angle, about $30^\circ$.
Different explanations are possible for such a conical structure,
such as precession of a jet (Iaria et al. 2008).
  However, this appears less likely
because the axis of the jet did not
change in $10$ years (Tudose et al. 2008). In our understanding this
neutron star is a good candidate for conical outflows, because (1)
it has episodes of very low and very high accretion rates, which is
very favorable for formation of conical outflows, (2) a neutron star
has only weak magnetic field which can be strongly compressed almost
to the stellar surface by the disk which is favorable for conical
outflows. This also means that velocity of outflows are of the
order of the Keplerian velocity at the surface of the neutron star.

\subsection{Application to black hole hosting systems}

Jets and winds are observed from accreting black holes  including both,
stellar-mass black holes and black holes in galactic nuclei.
The correlation between enhanced accretion rate and outflows has been extensively discussed
and observational data are in favor of this correlation.
  On the other hand the possibility of  magnetic flux accumulation
in the inner disk around the black hole and
the enhancement it gives to jets has been discussed and  observed in numerical simulations (Lovelace et al. 1994; Meier 2005; Igumenshchev 2008).
In this situation periods of
enhancement of the accretion rate may also lead to bunching of the field lines into an
X-type configuration (near the inner radius of the disk) and to formation of conical winds, because the mechanism
of conical winds does not require the dipole or any other special configuration.
Similarly, the magnetic field of a star may be more complex than a dipole one (see e.g., Donati et al. 2006; Mohanty and Shu 2008).

\begin{figure*}
\centering
\includegraphics[width=6in]{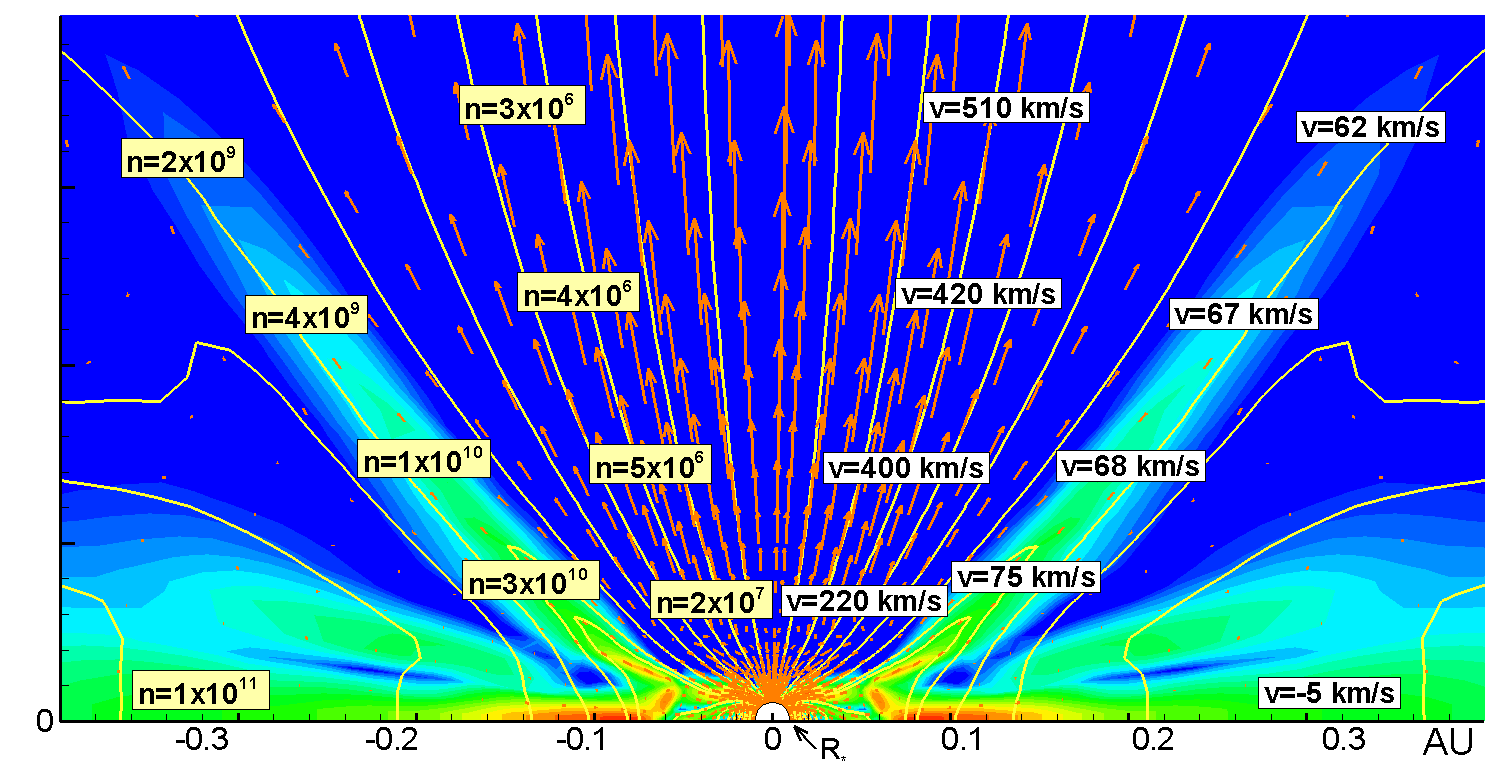}
\caption{Outflows in the propeller regime. The background shows
matter flux, lines are selected field lines, arrows are proportional
to velocity. Labels show total velocity and density at sample
points.}\label{prop-numb}
\end{figure*}

\begin{figure*}
\centering
\includegraphics[width=6in]{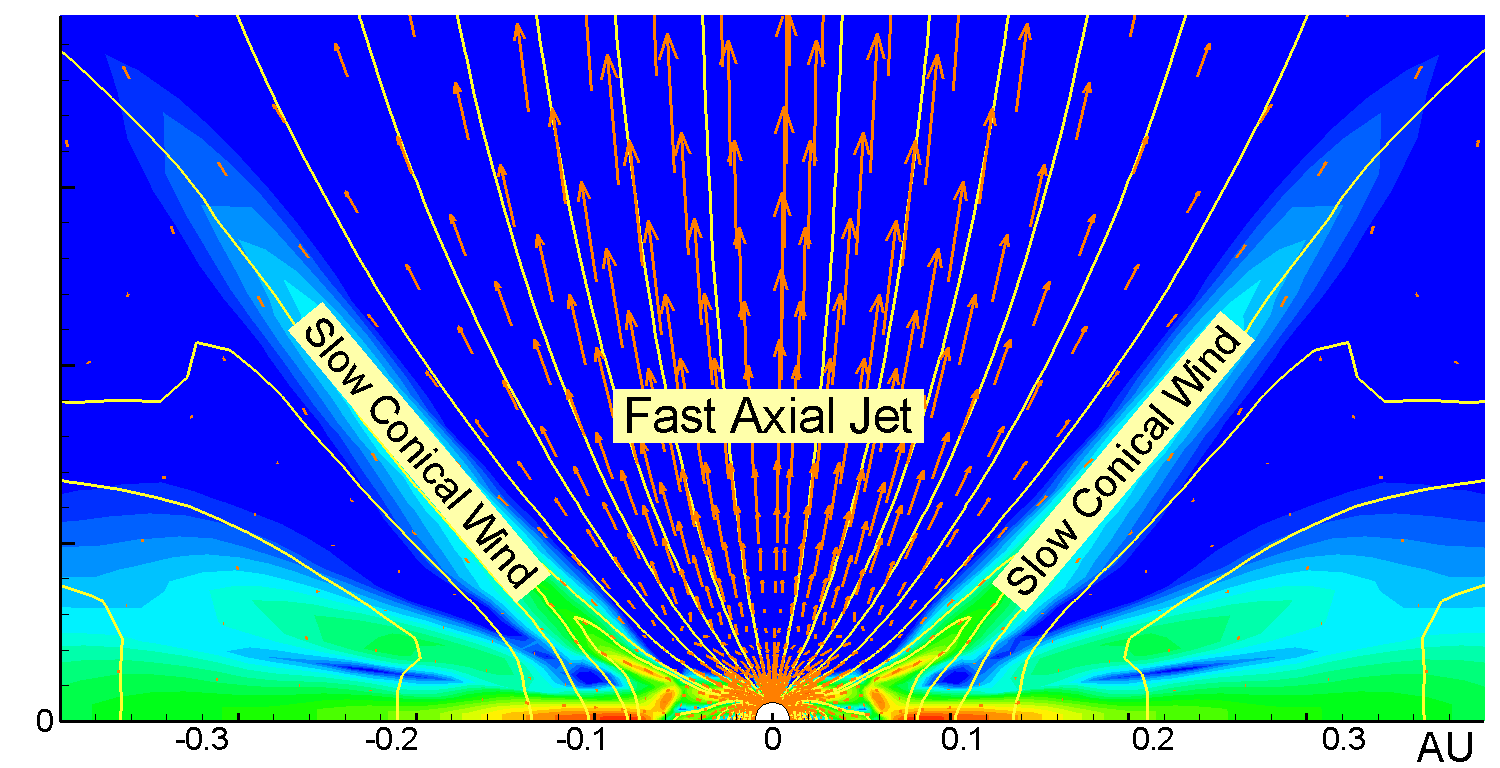}
\caption{Two components of outflows in the propeller regime.}\label{prop-label}
\end{figure*}

\begin{figure*}
\centering
\includegraphics[width=6in]{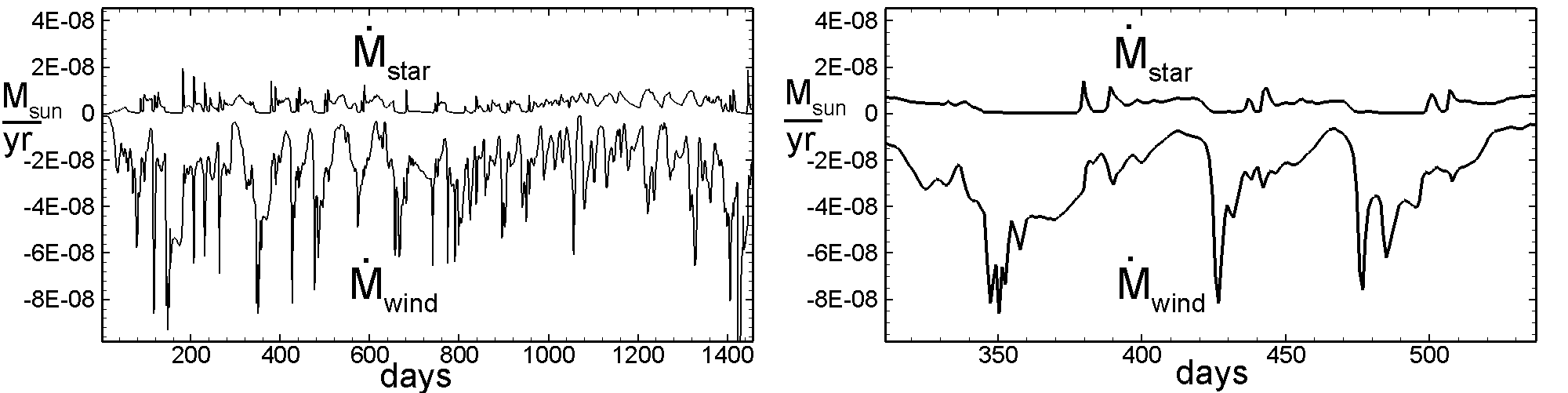}
\caption{Two components of outflows in the propeller regime.}\label{flux-prop}
\end{figure*}

\begin{figure*}
\centering
\includegraphics[width=6in]{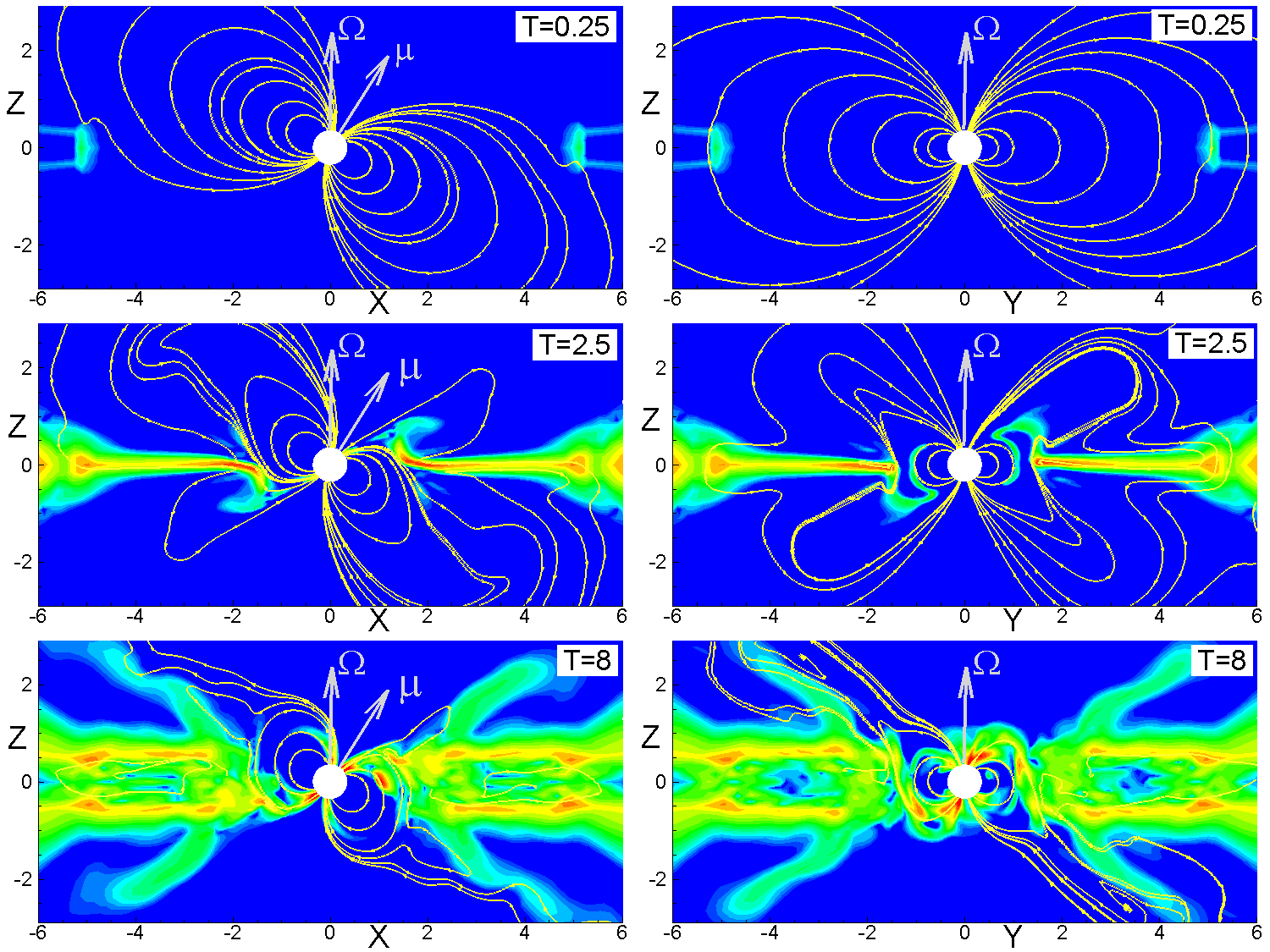}
\caption{Projections show formation of conical winds obtained in 3D
MHD simulations. Left panels show an XZ slice in the distribution of
matter flux (background) and magnetic field lines in different
moments of time. Right panels show corresponding YZ slices. Arrows
show the direction of the magnetic moment of the star $\mu$ and
angular velocity of rotation $\Omega$.}\label{3d-6}
\end{figure*}

\section{Outflows in the ``Propeller" Regime}

In case of accreting magnetized stars outflows may be associated
with ``propeller" stage of evolution (e.g., Illarionov \& Sunyaev 1975; Alpar \& Shaham 1985; Lovelace et al. 1999) if a star rotates sufficiently
fast. At propeller stage fast rotating magnetosphere transfers its
angular momentum to the disk matter (e.g. Lovelace et al. 1999).
Simulations show that most of matter flows from the inner parts of
the disk to widely open cones (Romanova et al. 2005; Ustyugova et
al. 2006). The process is quasi-periodic, or episodic (with no
definite period) which is connected with opening and closing of
magnetic field lines connecting a star and the disk. Such
quasi-periodic reconstruction of the magnetosphere due to
inflation and reconnection has been predicted theoretically (Aly
\& Kuijpers 1990, Uzdensky, Litwin \& K\"onigl 2003) and has been
observed in a number of axisymmetrtic simulations (Hirose et al.
1997; Goodson et al. 1997, 1999; Matt et al. 2002; Romanova et al.
2002; von Rekowski \& Brandenburg 2004). Multiple events of inflation
and reconnection (hundreds of outbursts) were observed in
simulations of the propeller stage (Romanova et al. 2005;
Ustyugova et al. 2006; see animation 
at http://www.astro.cornell.edu/$\sim$romanova/propeller.htm).

In the propeller regime the magnetosphere rotates faster than inner
region of the disk. This occurs if the co-rotation radius
$R_{cr}=(GM/\Omega_*^2)^{1/3}$ is smaller than magnetospheric radius
$R_m$ (e.g., Lovelace et al. 1999). Young stars are expected to be
in the propeller regime in two situations: (1) At the early stages
of evolution (say, at $T < 10^6$ years), when the star formed but
did not have time to spin-down, and (2) at later stages of
evolution, such as at CTTS stage, when the star is expected to be on
average in the rotational equilibrium state (e.g., Long et al. 2005)
but variation of the accretion rate leads to variation of $R_m$
around $R_{cr}$, where $R_{cr} < R_m$ is possible. We performed
axisymmetric simulations of accretion to a star in the propeller
regime, taking a star with the same parameters as in case of conical
winds, but with period $P_*=1$ day (Romanova et al. 2005; Ustyugova
et al. 2006). We chose $\alpha_v=0.3$ and $\alpha_d=0.1$ and thus
bunched the field lines to the X-type configuration  We observed
that in addition to conical wind there is a fast axial jet (see \fig{prop-numb})
so that the outflow has two components (see \fig{prop-label}).
The conical wind in this case is much more powerful - it carries most of the disk
matter away. The axial jet carries less mass, but it is accelerated
to high velocities. Acceleration occurs due to the magnetic pressure
of the ``magnetic tower" which forms above the star as a result of
winding of magnetic field lines of the star. Outbursts to conical
winds occur sporadically with a long time-scale interval (see \fig{flux-prop})
which is connected with the long time-scale interval of
accumulation and diffusion of the disk matter through the
magnetosphere of the star (see also Goodson et al. 1997; Fendt
2008). These propeller outflows were obtained in conditions
favorable for such a process: when the star rotated fast and an
X-type configuration developed. Future simulations should be done
for the case of propeller-driven outflows from slower rotating CTTS.
Collimation of conical winds may occur at larger distances from the
star for example, by disk winds (e.g., K\"onigl \& Pudritz 2000;
Ferreira et al. 2006; Matsakos et al. 2008).

\section{3D Simulations of Conical Winds}

We performed exploratory simulations of conical winds in global 3D
simulations.   One of main questions is what is an expected
direction of conical winds when magnetosphere of the star is
misaligned relative to the rotational axis. We chosen a case when
magnetosphere of a star is misaligned relative to the rotational
axis at an angle $\Theta=30^\circ$. We took a cubed sphere code
developed earlier (Koldoba et al. 2002, Romanova et al. 2003) and
used a grid $N_r\times N^2= 120\times 51^2$ in each of 6 blocks of
the sphere.  At such fine grid we were able to handle the case with
the low-density corona $\rho_c=0.001$ and took the density in the
disk $\rho_d=2$ which is 5 times lower than in the axisymmetric
case. However we also took smaller magnetic moment of the star,
$\tilde\mu=2$ (compared to $\tilde\mu=10$ in the axisymmetric case),
again, to save computing time. We did not push a disk from too large
distances (to save computing time) but from $r=5$ which is far
enough to insure an initial bunching of the magnetic field lines.
Subsequently, the bunching had been supported by sufficiently high
viscosity, $\alpha_{vis}=0.3$. We do not have diffusivity in the 3D
code however at such a grid estimated numerical diffusivity at the
disk-magnetosphere boundary is at the level $\alpha_d\sim 0.01-0.02$
so that conditions for conical winds are satisfied, that is $Pr_m >
1$.

Simulations have shown that accreting matter bunched field lines and
that some matter went to conical winds.  \fig{3d-6} shows that
conical winds are geometrically symmetric relative to the rotational
axis of the star $\Omega$ in the sense that an angle between
different parts of the conical winds and $\Omega$ axis is
approximately the same. However the density distribution in the wind
shows a spiral structures in the form of density amplification,
which rotates with an angular velocity of the star, $\Omega_*$, and
represents a one-arm spiral from each side of the outflow.

Note that at $\alpha_{vis}=0.3$ the disk-magnetosphere interaction
has tendency to be unstable relative to the interchange instability
(Romanova, Kulkarni \& Lovelace 2008; Kulkarni \& Romanova 2008) and
we do observe partial accretion through instabilities in addition to
funnel stream accretion which dominates at pretty high inclination
of the dipole $\Theta=30^\circ$ (Kulkarni \& Romanova 2009). One can
see that conical winds start blowing from those parts of the disk
which are quite far away from the disk-magnetosphere boundary. We
can guess that again, the accumulation of the $B_\phi$ component
above the disk drives matter outward to the conical winds. However
we should be caution in giving global conclusion about
non-dependence of conical winds from instabilities. Penetration of
the inner disk matter through the magnetosphere through
instabilities decreases the density at the inner edge of the disk
and ability of the disk matter to strongly compress the
magnetosphere which is a favorable condition for strong conical
winds. This possible interrelation between instabilities and conical
outflows  should be checked in the future 3D simulations in greater
detail.

\section{Investigation of parameter range}

To investigate the dependence on different parameters we took the main
case and varied one parameter at a time.
  The main effort has been
directed at understanding the dependence on the
effective viscosity and the effective magnetic
diffusivity. We performed two sets of runs: (1) at fixed diffusivity
and different viscosity coefficients, and (2) at fixed viscosity and
different diffusivity coefficients.

\smallskip

\noindent{\bf Dependence on viscosity at fixed
diffusivity.}
We fixed the diffusivity at $\alpha_{dif}=0.1$ and varied the viscosity
coefficient in
the range $\alpha_{vis} = 0.01 - 1$. We observed that for small
viscosity, $\alpha_{vis} < 0.1$, the magnetic field of the dipole
diffuses through the inner regions of the incoming disk and an X-type
configuration does not form. No conical outflows appear in this
case. We conclude that formation of conical winds requires
$\alpha_{vis} > \alpha_{dif}$, that is, $Pr_m > 1$.
   Next
we increased $\alpha_{vis}$ and observed that the X-type
configuration formed and conical outflows were generated. We
observed that the accretion rate to the star strongly increases with
$\alpha_{vis}$, while outflow rate to the conical winds
increases but only slowly.
  For $\alpha_{vis}=0.1$, the matter fluxes to the star and to the wind are small
and approximately equal.  For $\alpha_{vis}=0.3 \& 0.4$, the wind
carries about $30\%$ and $20\%$ of mass correspondingly).
 Angular momentum carried to the star also strongly increases with
$\alpha_{vis}$. In all cases the star spins-up, because the magnetospheric
radius, $r_m\approx 1.2$ is smaller than corotation radius which is
$r_{cor}=3$.   That is, incoming matter brings positive angular momentum
to the star. Conical outflows carry angular momentum away from the
disk. We should remember however that it is only a small part of the total angular momentum
of the disk as  shown above.   For $\alpha_{vis}=0.3$ and $0.4$,
matter outflows to the conical winds in oscillatory fashion. We often
observe that  quasi-periodic oscillations of the accretion occur for
$Pr_m\approx 3-4$.
\smallskip

\noindent{\bf Dependence on diffusivity at fixed
viscosity.}
In the next set of runs we fixed the viscosity at $\alpha_{vis}=0.3$ and
varied the diffusivity: from $0.01 to 1$. Again, no conical winds were
formed for $\alpha_{dif} > \alpha_{vis}$.  At relatively high diffusivity,
$\alpha_{dif}=0.1,~ 0.3$, about $30\%$ of the incoming matter flows to the conical winds.  For $\alpha_{dif}=0.1$ ($Pr_m=3$) conical outflows oscillate,
while at $\alpha_{dif}=0.3$ ($Pr_m=1$) no oscillations are
observed. For very small diffusivity, $\alpha_{dif}=0.01$, the
conical outflows also form with slightly smaller matter flux to the
winds.  However, the accretion rate to the star is larger. This is
somewhat unexpected because one would expect that at small
$\alpha_{dif}$ matter of the disk would be trapped by strong
magnetic flux accumulated at the inner edge of the disk. The
possible explanation may be  that penetration of matter
of the disk through the disk-magnetosphere boundary is determined by
the local Reynolds number $Re_m=\Delta r v_r/\eta_m$, which is
determined not only by the diffusivity coefficient $\eta_m\sim
\alpha_{dif}$, but also by the velocity of the flow, $v_r$ which become
very small at the disk-magnetosphere boundary, and also by the size
$\Delta_r$ of the boundary, which also become very small at high
Prandtl numbers.  Angular momentum fluxes to the star and to the
winds are approximately the same, excluding the case $\alpha_{dif}=0.01$ where the flux to the star is
larger. This
approximate equality {\it does not} mean that the conical winds
carry angular momentum flux out of the star. Instead, they carry
angular momentum from the inner regions of the disk. The remaining
angular momentum of the disk goes to the
spinning-up the star and part of the angular momentum of the
disk flows outward in the disk due to the viscous stress.

\smallskip

\noindent{\bf Variation of other parameters.} We varied the {\it period of the star (corotation radius)} taking
$r_{cor} = 5, ~10$ for slowly rotating stars and $r_{cor}=1.5,~ 2$ -
for faster rotating stars. In the case of slowly rotating stars conical
winds form and the outflow rate to
the winds is similar to one in the main case for
$r_{cor}=3$, though accretion rate to the star is somewhat larger.
For faster rotating star, $r_{cor}=2$, the amplitude of variability
increased and matter flux to outflows increased up to $50\%-70\%$
compared to the accretion rate to the star.  For even faster
rotation, $r_{cor}=1.5$, the accretion rate decreased  by a
factor of $5$ compared
to the main case, while the outflow to conical winds increased
by a factor of $2$.
Thus most of matter is ejected to conical winds. This situation
become close to the propeller regime, where the corotation radius becomes
close to the magnetospheric radius, $r_m\approx 1.2$ where most of
the incoming matter
may be ejected to the outflows (Lovelace et al. 1999; Romanova et al. 2005; Ustyugova et al. 2006).

Outflow of matter to winds occurs if the corona is not very
dense so that outflowing matter of the winds does not lose its
energy while propagating through the corona. In the main simulation
runs an initial density of the corona is $10^{-4}$ times the disk
density ($\rho_d=10$ versus $\rho_c=10^{-3}$). To test the
dependence on the coronal density we decreased its density by a factor of
$3$ and first chose smaller (compared to the main
case) transport coefficients, $\alpha_{vis}=0.1$,
$\alpha_{dif}=0.03$, hoping to enhance outflows to the winds.
Simulations have shown that matter fluxes to the star and to the winds are not appreciably different from the main case. We conclude that the coronal density used in the main case is sufficiently small as to not
 suppress  the outflows.
\smallskip

\noindent{\bf Summarizing the above subsections we conclude that:}

\smallskip
\noindent{\bf (1) Strong conical outflows} appear at a wide range of parameters
if both the magnetic viscosity and diffusivity are not very small, $\alpha_{\rm
v} > 0.1$ and $\alpha_{\rm d} > 0.01$. Outflows are most powerful
when the viscosity is a few times larger than diffusivity. For these
parameters the viscosity is high enough to drive disk matter inward to
the region of stronger magnetic field, while the diffusivity is high
enough to ensure the penetration of disk matter through the bunched
field lines.

\smallskip

\noindent{\bf (2) Weak outflows} are observed when the viscosity is
large and the diffusivity is very small. In this case the field lines
are bunched into the X-type configuration. However, the diffusivity
is not high enough to ensure loading of the disk matter to the field
lines of the magnetosphere.

\smallskip

\noindent{\bf (3) Slow accretion, no outflows} occurs when the viscosity
is small, $\alpha_{\rm v} < 0.1$ for any value of the diffusivity $\alpha_{\rm
d}$ . In this case the disk is stopped at larger distances from the
star by the strong magnetic field of the star.  Only very week
outflows or no outflows are observed in this case.

\section{Conclusions}

We discovered a new type of outflows - hollow conical winds - in MHD 
simulations of disk accretion to a rotating
magnetized star.
The conical winds occur under conditions where the poloidal magnetic 
field is bunched into an X-type
configuration.
\smallskip

\noindent{\bf In some respects these winds
 are similar} to
the X-winds
proposed by Shu and collaborators (e.g., Shu et al. 1994):

\smallskip

\noindent {\bf 1.} They
both require bunching of the field lines;

\smallskip

\noindent {\bf 2.} They both have high
rotation of the order of Keplerian rotation at the base of outflow,
and gradual poloidal acceleration;

\smallskip

\noindent {\bf 3.} They both are driven by the magnetic force
(see also Lovelace et al. 1991).

\medskip

\noindent {\bf However, there are a number of important
differences}:
\smallskip

\noindent {\bf 1.} Conical winds flow in a thin shell (they are hollow), while X-winds
flow at different angles  below the ``dead zone";

\smallskip

\noindent {\bf 2.} Conical winds
form around stars of any rotation rate including slow rotation, and
do not require the fine tuning of angular velocity of the inner disk
to that of magnetosphere;

\smallskip

\noindent {\bf 3.} Conical winds are non-stationary: the
magnetic field constantly inflates and reconnects;

\smallskip

\noindent {\bf 4.} Conical winds
carry away part of the angular momentum of the inner disk and are
not responsible for spinning-down the star, while X-winds are
predicted to take away angular momentum from the star;

\smallskip

\noindent {\bf 5.} In the conical winds there is a
fast component of the flow along field lines threading the star.

\smallskip

\noindent {\bf 6.} In the propeller regime (where
the corotation radius is comparable or
smaller than magnetospheric radius) a new component appears: a strong 
magnetically dominated  jet along the open field lines of the
star.  In this component
low density matter is accelerated up
to super-Keplerian velocities.  The star loses its angular momentum mainly 
due to the twisted magnetic
field of this component.

\noindent
Some of these differences, such as non-stationarity of conical winds is 
connected with natural restrictions of the stationary model of X-winds.

Conical winds can explain the hollow conical shape of outflows near
young stars of different type (CTTSs, EXors, Type I objects) which
have been recently resolved.
The conical winds may arise
from any type of the disk-accreting magnetized
stars, including accreting young stars, white dwarfs, neutron stars
and possibly black holes as mentioned.
 An exact configuration of the
magnetic flux is not important: only the bunching/inclination and inflation
of the flux are important. This makes this mechanism universal
and it may work in different situations.
It is often observed that enhancement of the accretion rates leads to enhanced
outflows. Mechanism of conical winds gives natural explanation of such events observed
in a wide variety of astrophysical objects.

It is still not known what are the values of transport coefficients in real accretion disks.
MRI simulations have shown that magneto-rotational mechanism of
turbulent transport may give values of $\alpha$ in the range: $\alpha_{vis} = 10^{-2} -
0.4$ (e.g., Stone et al. 2000) and corresponding diffusivity values.
If in
real disks transport coefficients are large, $\alpha\sim 0.1$ then strong outflows
are expected during periods when $\alpha_{vis} > \alpha_{dif}$ with steady flow 
to the wind or with short-period variability like in \fig{fluxes-4}. If in the 
opposite case the transport coefficients are
always small, say
$\alpha_{vis}=\alpha_{dif}=0.01$, then conical outflows are expected 
only during periods of enhanced accretion when accretion rate is 
enhanced due to e.g. some
instabilities or other mechanism of angular momentum transport, 
e.g., Rossby waves
(Lovelace et al. 1999; Li et al. 2000). Then outflows will be 
characterized by  high-amplitude outbursts
to conical winds with a large time-interval between outbursts 
determined by the small diffusivity
at the disk-magnetosphere boundary (similar to that
in the propeller regime (e.g., \fig{flux-prop}).

\smallskip

\noindent{\bf If a star is in the propeller regime, then clear two-component 
outflows are observed:}
\smallskip

\noindent (1) Conical-type winds which may carry away significant part of 
the {\it disk mass and angular momentum}, with velocities approximately 
equal to Keplerian velocity at the base of the winds.
\smallskip

\noindent (2) Magnetically dominated axial jet where low-density matter
is accelerated near the star to much higher velocities.
The twisted magnetic field of this jet carries away most
of the angular momentum of the star.
\smallskip

 The propeller regime may be responsible for the fast loss of angular momentum by young stars
before their CTTS stage. It may also appear in stars which are in rotational equilibrium
during periods of the  accretion rate is lower.
\smallskip

The propeller-driven axial jet has  good collimation due to interaction
with surrounding conical wind.
  The overall conical wind may be collimated at larger distances by
disk winds  emanating from the disk at large radii
(Blandford \& Payne 1982) or the pressure of the external
medium (Lovelace et al. 1991).
 Typical terminal velocity of outflows
from the disk are of the order of the Keplerian velocity at the
radius from which matter flows to the wind and is expected to be
smaller than velocity in conical winds, though the total angular
momentum may  be larger (e.g., Ferreira et al. 2006).

\section*{Acknowledgments}

The authors thank S. Edwards, S. Cabrit, E. Dougatos, J. Ferreira and F. Shu  
for helpful discussions.
The authors were supported in part by NASA grant NNX08AH25G and by
NSF grants AST-0607135 and AST-0807129. MMR thanks NASA for use of
the NASA High Performance Facilities. AVK and GVU were supported in
part by grant RFBR 06-02016608, Program 4 of RAS.

\end{document}